\documentclass{emulateapj}
\usepackage{apjfonts}

\newcommand{\deltavec}{\mbox{\boldmath $\delta$}}

\lefthead{HAN} 
\righthead{PLANET VS BINARY}

\begin{document}
\title{Distinguishing between Planetary and Binary Interpretations of Microlensing
Central Perturbations under Severe Finite-Source Effect}

\author{Cheongho Han}
\affil{Program of Brain Korea 21, Department of Physics, 
Chungbuk National University, Chongju 361-763, Korea;
cheongho@astroph.chungbuk.ac.kr}

\submitted{Submitted to The Astrophysical Journal}

\begin{abstract}
In the current strategy of microlensing planet searches focusing 
on high-magnification events, wide and close binaries pose 
important sources of contamination that imitates planetary 
signals.  For the purpose of finding systematic differences, 
we compare the patterns of central perturbations induced by 
a planet and a binary companion under severe finite-source 
effect.  We find that the most prominent difference shows up 
in the morphology of the edge features with negative excess 
that appear at the edge of the circle with its center located 
at the caustic center and a radius equivalent to the source 
radius.  It is found that the feature induced a binary companion 
forms a complete annulus, while the feature induced by a planet 
appears as several arc segments.  This difference provides a 
useful diagnostic for immediate discrimination of a planet-induced 
perturbation from that induced by a binary companion,   
where the absence of a well-developed dip in the residual 
from the single-lensing light curve at both or either of the moments of the 
caustic center's entrance into and exit from the source 
star surface indicates that the perturbation is 
produced by a planetary companion.
We find that that
this difference is basically caused by the difference between 
the shapes of the central caustics induced by the two different 
types of companions.
\end{abstract}



\section{Introduction}

Microlensing planets are being discovered at an accelerating rate 
and the total number of detections now reaches 8 \citep{bond04, 
udalski05, beaulieu06, gould06, gaudi08, dong08, bennett08}.  
The microlensing signal of a planet is a brief perturbation to 
the smooth standard light curve of the primary-induced lensing 
event occurring on a background star. To achieve the monitoring 
frequency required to detect short-duration planetary signals, 
current planetary lensing searches are being conducted by using 
a combination of survey and follow-up observations, where alerts 
of ongoing lensing events are issued by the survey observations 
(OGLE: Udalski et al.\ 1994, MOA: Bond et al.\ 2001) and the 
alerted events are intensively monitored by  the follow-up 
observations (Micro-FUN: Dong et al.\ 2006-, PLANET: Kubas et 
al.\ 2008).  However, the number of telescopes available for 
follow-up observations is far less for intensive monitoring of 
all alerted events \citep{dominik08} and thus observations are 
focused on events which can maximize the planet detection 
probability with follow-up observations. Currently, the prime 
targets of follow-up observations are high-magnification events 
for which the source trajectories always pass close to the 
perturbation region around the central caustic induced by a 
planet and thus planet detection efficiency is intrinsically 
high \citep{griest98}.

Although high-magnification events yield high sensitivity to 
planets, interpretation of the observed perturbation often
suffers from several potential degeneracies.  There are two 
major degeneracies causing this complication.  The first 
well-known wide/close degeneracy arises due to the fact that 
the perturbation induced by a planet with a projected separation 
in units of the Einstein ring, $s$, is very similar to the 
perturbation induced by a planet with a separation $1/s$ 
\citep{dominik99, an05, chung05}.\footnote{We note that the 
wide/close degeneracy occurs not only for planetary events
but also for binary events in general \citep{dominik96}.}  
The other degeneracy arises due to the fact that a central 
perturbation of a high-magnification event can also be produced 
by a very close ($s\ll 1$) or a very wide ($s\gg 1$) binary with 
roughly equal mass components. Hereafter, we refer the latter 
degeneracy as the `planet/binary' degeneracy. In the sense that 
the planet/binary degeneracy causes indeterminacy of both of 
the mass ratio and the separation between the lens components, 
while the wide/close degeneracy causes ambiguity only in the 
separation, the planet/binary degeneracy poses more serious 
problem in the interpretation of an observed perturbation. 
Fortunately, the perturbations induced by a planet and a binary 
are intrinsically different and thus it would be possible to 
determine whether the perturbation is caused by a planet or a 
binary. However, distinguishing between the two interpretations 
usually requires detailed modelling which demands time-consuming 
search for a solution in the vast space of many parameters.  
Therefore, a simple diagnostic that can resolve the planet/binary 
degeneracy would be very helpful not only for the prompt interpretation 
of an observed perturbation but also for the establishment of the 
observational setup  optimizing the coverage of the characteristic 
features helping to resolve the degeneracy.  \citet{han08a} provided 
such a diagnostic but this diagnostic applies to a subset of light 
curves with double-peak features.

Recently, \cite{dong08} reported a planet detected  from the 
analysis of a new type of high-magnification event where the 
angular extent of the perturbation region induced by the planet 
is significantly smaller than the angular size of the source 
and thus finite-source effect is very severe.  The main 
perturbation features of this event are double spikes in the 
residuals, which are approximately centered at the times when 
the lens enters and exits the source.  In 2008 season, several 
more such events were detected (A.\ Gould 2008, P.\ Fouque 
2008, private communication), implying that events with these 
features might be common.  From detailed investigation, 
\citet{han08b} found that spike features commonly appear for 
planetary-lensing events affected by severe finite-source effect 
and thus they can be used for the diagnosis of the existence 
of a planet. However, they noted that such features can also 
be produced by a wide or a close binary companion and thus the 
existence of the feature does not necessarily confirm the 
planetary interpretation of the perturbation.

In this paper, we investigate systematic differences between 
the patterns of central perturbation induced by a planet and 
a binary companion.  From this,  we identify a diagnostic that 
can be used to immediately distinguish between the planetary 
and binary interpretations. We also investigate the origin of 
the difference.

\section{Perturbation Pattern}

The pattern of central perturbations is basically determined 
by the caustic shape.  The shapes of the caustics induced by a 
planet and a binary companion are intrinsically different and 
thus the resulting patterns of the perturbations produced by 
the two types of lens systems will be different.

For a wide binary system, the central caustic has a hypocycloid 
shape with four cusps.  When the horizontal and vertical widths 
of the caustic are measured as the separations between the two 
confronting cusps located on and off the binary axis, respectively, 
the ratio of the vertical to horizontal widths is represented by 
\citep{han05}
\begin{equation}
R_b\sim \left( {1-\gamma\over 1+\gamma}\right)^{1/2};\qquad
\gamma={q\over (1+q)s^2},
\label{eq1}
\end{equation}
where $\gamma$ represents the shear exerted by the companion.
As the binary separation increases, the shear decreases and 
the width ratio becomes $R_b\rightarrow 1$, implying that the 
horizontal and the vertical widths are nearly identical for a 
wide binary.  In addition, the shape of the caustic is symmetric 
with respect to both of the binary axis and the axis normal to 
the binary axis.  The caustic induced by a close binary is 
approximately identical to that of the wide binary with a 
separation of $s^{-1}$.

In contrast to the central caustic induced by a binary 
companion, the central caustic induced by a planet has an 
elongated wedge-like shape.  The width ratio of the planetary 
central caustic defined in a similar way as that of the binary 
case is represented by \citep{chung05}
\begin{equation}
R_p \sim {(s-s^{-1})|\sin^3\phi| \over
(s+s^{-1}-2\cos\phi)^2},
\label{eq2}
\end{equation}
where 
$\cos\phi=(3/4)(s+s^{-1}) \{ 1-[32/9(s+s^{-1})^2]^{1/2}\}$.
For the range of planetary separations where the size of 
the central caustic is not negligible (see \citet{han09}), 
the width ratio of the central caustic is substantially 
smaller than unity, implying that the caustic is elongated 
along the planet-primary axis.  In addition, three of the 
four cusps of the caustic lean toward the primary direction.  
As a result, the caustic is {\it not} symmetric with respect 
to the axis normal to the planet-primary axis.

Another important factor that affects the perturbation pattern
is the effect of the finite size of a source star.  The lensing 
magnification affected by the finite-source effect corresponds 
to the magnification averaged over the source star surface.
As a result, the signal of the companion is smeared out by the 
finite-source effect and the degree of the effect depends on 
the ratio of the caustic size to the size of the source star.

\subsection{Excess Maps}

To see the difference between the patterns of central perturbations 
induced by a planet and a binary companion under severe finite-source 
effect, we construct maps of magnification excess around the central 
caustics of the individual lens systems.  The magnification excess 
is defined as
\begin{equation}
\epsilon={A-A_0\over A_0},
\label{eq3}
\end{equation}
where $A$ and $A_0$ represent the lensing magnifications with 
and without the companion, respectively.  For the computation 
of the magnification, we use a ray-shooting based algorithm 
developed by \citet{dong06}.  The algorithm is optimized for 
high-magnification events and saves computation time by limiting 
the range of ray shooting on the image plane to a narrow annulus 
around the Einstein ring.  We take the finite-source effect into 
consideration by modelling the source brightness profile as 
\begin{equation}
{I(\theta)\over I_0}=
1-\Gamma\left( 1-{3\over 2}\cos\theta\right)
-\Lambda \left( 1-{5\over 4}\cos^{1/2}\theta\right),
\label{eq4}
\end{equation}
where $\theta$ is the angle between the normal direction to the 
source-star surface and the line of sight.  We adopt a linear 
and a square-root limb-darkening coefficients of $(\Gamma, 
\Lambda)=(-0.46,1.11)$.

\begin{figure}[t]
\epsscale{1.2}
\plotone{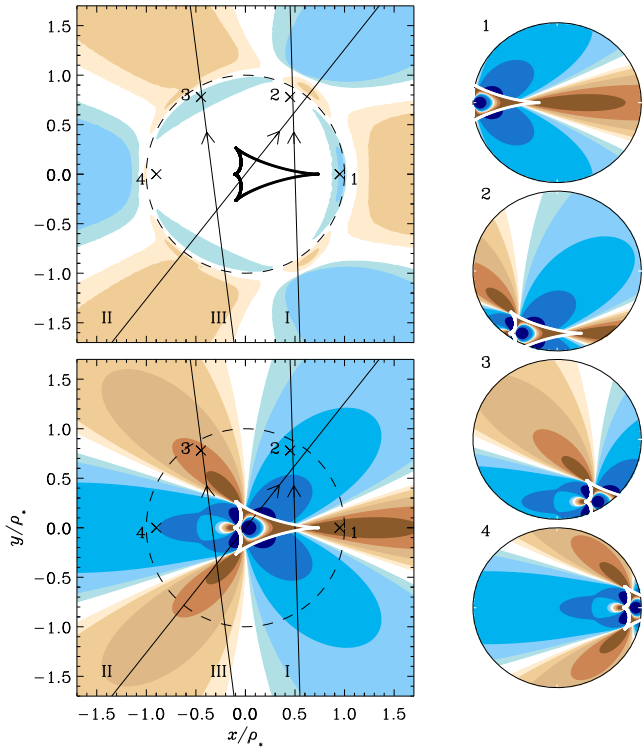}
\caption{\label{fig:one}
Color-scale maps of magnification excess around the central 
caustic induced by a planet.  The map in the upper left panel 
is constructed by considering the finite-source effect while 
the map in the lower left panel is for a point source.  The 
abscissa of the map is parallel with the axis connecting the 
two lens components and the center is located at the caustic 
center.  In each map, the planet is located on the right and 
all lengths are normalized by the source radius $\rho_*$.  
Colors are chosen such that the regions with brown and blue-tone 
colors represent the areas where the magnification is higher 
($\epsilon > 0$) and lower ($\epsilon < 0$) than the single-lensing 
magnification without the companion, respectively.  For each 
tone, the color scale becomes darker at the excess levels of 
$|\epsilon| =2\%$, 4\%, 8\%, 16\%, and 32\%, respectively.  
Each of the encircled maps on the right side shows the excess 
pattern enclosed by the source star at the time when the center 
of the source star is located at the position marked by `x' 
with a corresponding number.  The straight lines with arrows 
represent the source trajectories where the residual of the 
light curves of the resulting events from that of a single-lensing 
event are presented in Fig.~\ref{fig:three}.  
}\end{figure}

\begin{figure}[t]
\epsscale{1.2}
\plotone{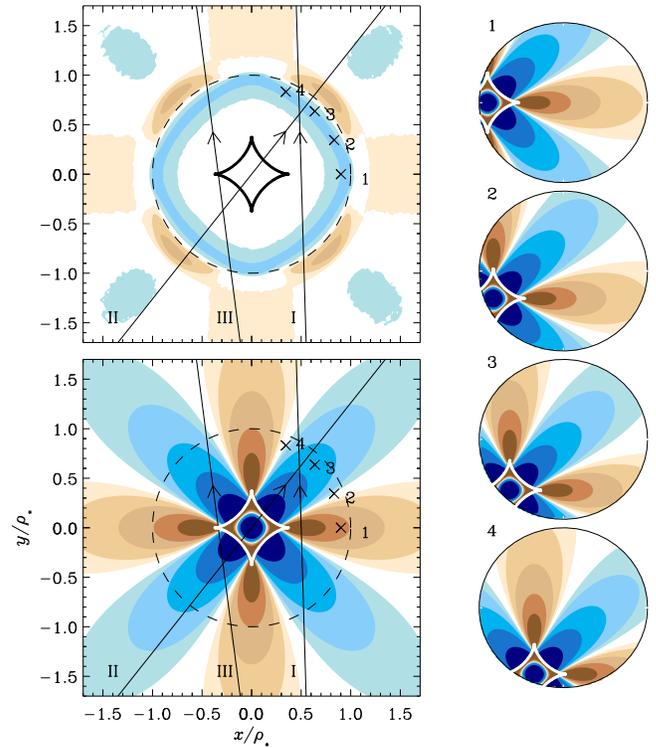}
\caption{\label{fig:two}
Color-scale maps of magnification excess around the central 
caustic induced by the companion of a wide-separation binary.
Notations are same in Fig.~\ref{fig:one}.
}\end{figure}

Figure~\ref{fig:one} and \ref{fig:two} show the constructed 
excess-pattern maps represented in color scales.  In each 
figure, the map in the upper left panel is constructed by 
considering the finite-source effect while the map in the 
lower left panel is for a point source.  The abscissa of 
the map is parallel with the axis connecting the two lens 
components and the center is located at the caustic center, 
that is located at the position with an offset from the position 
of the primary of 
\begin{equation}
\deltavec = 
\cases{
 {\bf s}^{-1}q/(1+q)    & for $s>1$, \cr
-{\bf s}[(1+q)^{-1}-1]  & for $s<1$,  \cr
}
\label{eq5}
\end{equation}
where the sign is positive when the offset vector is directed 
toward the companion and vise versa.  The companion is 
located on the right and all lengths are normalized by the 
source radius.  Colors are chosen such that the regions with 
brown and blue-tone colors represent the areas where the 
magnification is higher ($\epsilon > 0$) and lower ($\epsilon 
< 0$) than the single-lensing magnification without the companion, 
respectively.  For each tone, the color scale becomes darker at 
the excess levels of $|\epsilon|=2\%$, 4\%, 8\%, 16\%, and 32\%, 
respectively.

From the maps, one finds several features that commonly appear 
in the perturbation patterns of both planetary and binary cases.  
The first feature is a region of little excess inside a circle 
with its center located at the caustic center and a radius 
corresponding to that of the source star (dashed circle in each 
map).  The other feature is the perturbation regions that appear 
at the edge of the circle with either positive or negative excess.  
\citet{han08b} pointed out that the very small excess inside the 
circle is caused by the cancellation of the positive and negative 
excesses by the finite-source effect.  They also indicated that 
the feature at the edge of the circle (hereafter we refer this 
feature as `edge feature') is formed by the break of the balance 
between the positive and negative excesses due to a partial 
coverage of the strong excess region around the caustic by the 
source star.  These features in combination result in a distinctive 
signal of a companion in the residual of a light curve, that is 
characterized by a spike of either positive or negative excess at 
the moment when the center of the caustic enters or exits the source 
star and a flat residual region between the spikes.  See Fig.\ 3 of 
\citet{han08b}.

\subsection{Diagnosis}

Along with the similarities, we find that there also exist 
systematic differences between the patterns of the perturbations 
induced by a planet and a binary companion.  The most prominent 
difference shows up in the morphology of the edge features with 
{\it negative} excess.  It is found that the feature induced a 
binary companion forms a complete annulus, while the feature 
induced by a planet appears as several arc segments.
We find that this trend holds for nearly all planetary 
systems with mass ratios $q\lesssim 10^{-2}$ and separations 
where the planet detection efficiency is important.
See Fig.\ 1 of \citet{han09}.

To search for the origin of these differences, we produce 
additional maps showing the region enclosed by the source star 
at the time when the source star is located at various positions 
of the annulus where the negative edge features systematically 
occur.  These maps are presented on the right side of 
Figure~\ref{fig:one} and \ref{fig:two}.  From close examination 
of the perturbation pattern, it is found that the morphological 
differences between the planetary and binary cases are basically 
caused by the difference in the shape of the caustics. The shape 
of the caustic induced by the planet lacks symmetry.  As a result, 
the location of the source position where the overall excess is 
negative depends on the orientation of the source position with 
respect to the caustic.  For example, when the source is located 
at the position on the annulus toward the direction of the 
sharp-pointed cusp marked as `1', the number of strong-negative 
excess regions within the area encompassed by the source star 
is three (two just outside the fold caustics and the other 
inside of the caustic) while there exists only a single 
strong-positive excess region (the one extending from the cusp), 
resulting in overall negative excess.  By contrast, when the 
source is located at the opposite position marked as `4', there 
are equal numbers of positive and negative excess regions of 
three, respectively, and thus the resulting excess is very small.  
In contrast to the shape of the planet-induced caustic, the shape 
of the binary-induced caustic is more symmetric and thus the 
regions of negative excess appear in all regions along the annulus.

\begin{figure}[t]
\epsscale{1.2}
\plotone{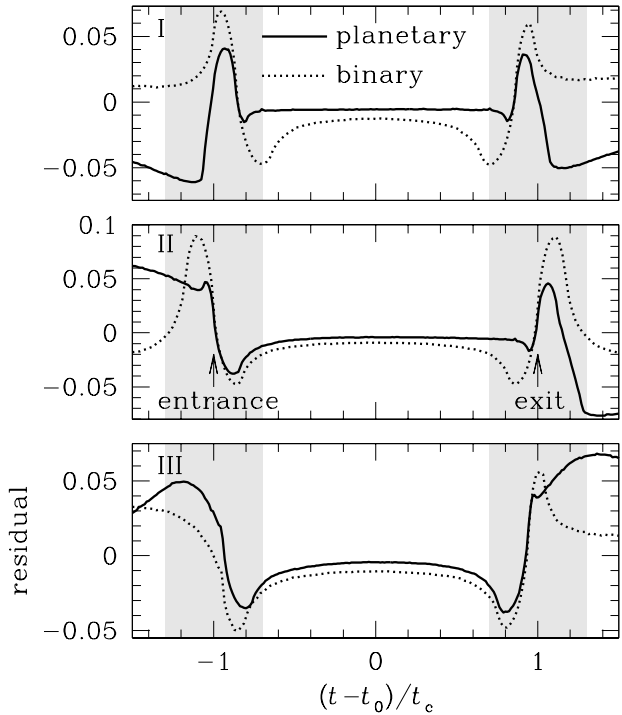}
\caption{\label{fig:three}
Residuals of lensing light curves from a single-lensing 
one for example planetary and wide-separation binary events.
Note that for the case of the binary, well-developed negative 
spikes occur at both moments when the center of the caustic 
enters and exits the source star.  On the other hand, a 
negative spike may {\it not} occur at one (middle panel) or 
both of the moments (upper panel). The source trajectories 
responsible for the individual events are marked in 
Fig.~\ref{fig:one} and \ref{fig:two} with the corresponding 
roman numbers.  Time is normalized by the duration required 
for the lens to cross the source radius, $t_c$.
}\end{figure}

The systematic difference in the morphology of the perturbation 
patterns enables one to distinguish between the perturbations 
caused by planetary and binary companions.  Due to the continuity 
of the negative-edge feature in the excess pattern, the perturbation 
in the light curve of a lensing event induced by a wide-separation 
binary will have well-developed negative spikes in the residual at 
both moments when the caustic center enters and exits the source 
star.  On the other hand, the edge feature induced by a planet is 
split into segments and thus the resulting light curve will not 
often have a negative spike in the residual at both or either of 
the moments of entrance and exit.  Therefore, the absence of the 
double negative-spike feature can be used as a simple diagnostic 
to immediately distinguish the planetary interpretation from the 
binary interpretation.

In Figure~\ref{fig:three}, we present residuals of lensing light 
curves of several example planetary (solid curve) and binary 
events (dotted curve), where the source trajectories responsible 
for the individual events are marked in Figure\ref{fig:one} and 
\ref{fig:two} with the corresponding roman numbers.  For the 
binary case, there always exist two well-developed negative 
spikes.  For a planetary case, on the other hand, there can be 
no (upper panel) or only a single (middle panel) spike.  Setting 
a critical value of the magnification excess to define the term 
"well-developed spike" is not easy due to the variation of the 
perturbation pattern combined with the difficulty in the conversion 
of the observed flux into magnification.  Nevertheless, the term 
can be defined in the observational point of view as {\it a 
well-resolved dip in the residual observed with high enough precision}.
Under this definition, a spike with deviation $\gtrsim 3\%$ can be 
well resolved considering that the photometric precision of the 
current follow-up observations reaches $\sim 1\%$ at the peaks of 
high-magnification events.  We note that a planet can produce 
perturbations with double negative-spike features as shown in the 
lower panel of Figure~\ref{fig:three}.  Therefore, the existence 
of double negative spikes does not necessarily confirm that the 
perturbation is caused by a binary companion.  In this case, the 
proposed diagnostic cannot be used and detailed modelling is 
required to distinguish between the two interpretations.

\section{Conclusion}

We compared the patterns of central perturbations induced by 
a planet and a binary companion under severe finite-source 
effect.  We found that the most prominent difference shows 
up in the morphology of the edge features with negative excess,
where the feature induced a binary companion forms a complete 
annulus, while the feature induced by a planet appears as 
several arc segments.  This difference provides a useful 
diagnostic for immediate discrimination of a planet-induced 
perturbation from that induced by a binary companion, where 
the absence of a well-developed dip in the residual from the 
single-lensing light curve at both or either of the moments 
of the caustic center's entrance into or exit from the source 
star surface indicates that the perturbation is produced by a 
planetary companion.  We found that that this difference is 
basically caused by the difference between  the shapes of the 
central caustics induced by the two different types of companions.

\acknowledgments 
This work was supported by the Science Research Center (SRC) 
program of Korea Science and Engineering Foundation (KOSEF).

\end{document}